# An Automated Petri-Net Based Approach for Change Management in Distributed Telemedicine Environment

S. Mtibaa and M. Tagina

**Abstract**— The worldwide healthcare industry is facing a number of daunting challenges which are forcing healthcare systems worldwide to adapt and transform, and will ultimately completely redefine the way they do business and deliver care for patients. In this paper, we present a distributed telemedicine environement reaping from both the benefits of Service Oriented Approach (SOA) and the strong telecoms capabilities. We propose an automated approach to handle changes in a distributed telemedicine environement. A combined Petri nets model to handle changes and Reconfigurable Petri nets model to react to these changes are used to fulfill telemedicine functional and non functional requirements.

——————————  ◆  ——————————

## 1 INTRODUCTION

INCREASING competition, financial pressure, and customer demands are causing service providers to transform the way they create and deliver services. The ultimate goal of this transformation is to achieve faster time-to-market, leverage shared enablers across applications, achieve capital and operational savings, and generate additional subscriber and non-subscriber revenues from non-traditional areas, such as advertising. Throughout this transformation, service providers are keeping an eye on the needs of end users [1].

Today's end users are looking for a consistent and seamless experience across diverse devices and networks. They want applications to work the same regardless of their location, who they are connected to, or what device they are using [2].

Telecom market research shows that end users are willing to pay a premium for services that can be personalized depending on the context within which they are being used (work, home, traveling, etc.) and they are willing to pay for services that allow them to blend capabilities or features that enhance their lifestyles and productivity [3]. In other words, consumers value the applications riding on the connection higher than the connection itself. Therefore, the value end users place on communication has shifted from basic broadband connectivity to services that enable contextaware, personalized and blended services. At the same time, web 2.0 is enabling collaboration and sharing among end users through social networking sites, wikis, blogs, and tags. Web 2.0 uses technologies such as Web Services to expose parts of applications that can then be combined and blended to create whole new services.

Obviously, end user preferences will continue to evolve, so a service provider's battle for differentiation must focus on the ability to foster rapid innovation (simplicity, flexibility and time-to-market) while providing quality of experience (QoE) for end users with service performance, quality, security, personalization, and usability [4].

Traditional service providers will not be able to keep up with the pace of service innovation taking place in the web 2.0 arena unless they engage in a services transformation process. That process must implement an effective Service Delivery Environment and impact factors that hinder their agility to develop, launch, manage and bill for services [6]. Service providers can achieve these goals and get a seat in the web 2.0 arena by collaborating with an ecosystem of partners. With the right partners, service providers can leverage partner capabilities in Web Services and allow controlled use of their own network services and capabilities through the most appropriate technologies to ensure a true QoE for end users. At the same time, they can create and transition to a service model that allows for the desired level of integration with the open Internet/web (2.0) world.

Modern healthcare is technology intensive, so dealing with complex technology challenges is not new. There has been significant progress in implementing internal and inter-organization IT solutions that provide better context for medical decisions reduce administrative costs and improve patient safety by reducing errors. Technical advances have included the development of IT standards for secure exchange of information, presentation of relevant patient information at a doctor's PC, and interworking of diverse imaging solutions. In this paper, we will propose a distributed telemedicine environment in form of virtual enterprise taking advantage of Web-based communications-centric and IT-centric capabilities. The virtual enterprise [5] is based on the ability to create temporary co-operations and to realize the value of a short business opportunity

- S. Mtibaa is with the LI3 Laboratory / University of Manouba, National School of Computer Sciences, 2010. E-mail: Sabri.Mtibaa@gmail.com.
- M. Tagina is with the LI3 Laboratory / University of Manouba, National School of Computer Sciences, 2010. E-mail: Moncef.Tagina@ensi.rnu.tn.

that the partners cannot (or can, but only to lesser extent) capture on their own. In this scope, we will concentre on potentiels of Web services to construct a full process from stem to stern with a special concentration on the business aspects. Moving a higher percentage of healthcare outside the clinic or hospital is a necessary and significant component of strategic change described. It is important to emphasize that is a part of that solution since the management of change to enable full view of the integrated lifecyle (how Web services are created, modified and disposed of) is a key aspect.

Change management is considered one of the challenging problems in the field of software engineering, database and workflow systems. In [6] [7], some frameworks dealing with the changes in tightly-coupled system have been proposed. A central monitoring module is usually deployed to handle, forward and self-adapt to the occurrence of changes. Due to known drawbacks of central approach, our work will not consider a central mechanism for change management.

As the composed/aggregated services available increases, the complexity of managing changes will grow and the manual management of changes become not pratical even impossible. Indeed, Web services technology cannot pretend to provide all answers in terms of patient safety and other medicine services but can make a significant contribution. One of the greatest promises of the distributed telemedicine environement is ability to self-adapting to garantuee goals acheiving. Abstracting changes in business relationships require an automated approach to manage changes without any impact on high level functionaltities offered to users. For example, the proposed telemedicine environement must be able to automatically plug-in/plugin-out Web services with little overhead while guaranteeing both functional and non-functional properties. Functional properties refer to the functionalities that telemedicine environement have to fulfill. The non-functional properties refer to events surrounding the functional properties. Change management is therefore a critical component in the deployment of telemedicine environement. Our works provides such as framework for change management based on a pre-defined use cases. To ensure desidered requirements, different pratical approach can be used. An example of these approaches is Buttom-up approach in which the changes are intiated in the Web service environement. As a case, a service operation may become unavailable during execution and trigger dependencies services and users in order to replace this service. In this paper, we will focus on this aspect.

Reasearchers interested to automate the change management process to make it transparent to users. A Petri net based approach to maintain correctness between process type and instances is presented [14]. Changes to process type occur when the process schema is modified in response to the environment. For example, a business process may adapt to comply with new legislation, or it may be optimized for performance reasons. In [15], a framework to detect and react to the exceptional changes that can be raised inside work ow-driven Web application is proposed. It first classifies these changes into behavioral (or user-generated), semantic (or application), and system exceptions. For example, the free user navigation through Web pages may result in the wrong invocation of the expired link, or double-click the link when only one click is respected.

We present an approach to automatic management of bottom-up changes using Petri nets. In our work, we use Petri nets to model handling and adptative changes in distributed telemedicine environement. Petri nets have been used to model a variety of concurrent and discrete event distributed systems [8]. We model changes using Petri nets because of their applicability to a Web services composition modeling. The behavior of a composite Web services is described by the evolution of its Petri net model. As the Petri net evolves, the system reaches different safe and unsafe states that can be completely defined by the marking of a Petri net model. Furthermore, Petri nets map directly to our change specification. They also preserve all the details of our change specification while modeling the changes accurately. For example, Petri nets can easily represent the safe an unsafe states of web services composition. They represent changes between these states as transitions. Moreover, the use of reconfigurable Petri nets allows us to incorporate our mapping rules into the Petri net model. This allows us to completely model our change specification, without the need to use additional modeling tools.

The remainder of this paper is organized as follows. In Section 2, we use the distributed architecture of telemedicine and a scenario from this domain to motivate our work. It will also be used as a running example. Section 3 presents a bottom-up specification of changes. In Section 4, we describe our change management model which is based on Petri nets. Finally, we conclude in section 5.

## 2 DISTRIBUTED TELEMEDICINE ENVINRONEMENT

In this section, we present the global context of our work and an overview about Service Oriented approach. A scenario from Telemedicine world will be exposed to prepare the introduction of changes model.

### 2.1 Context and Motivations

The Worldwide Web Consortium (W3C) defines a Web Service as a software system designed to support interoperable machine-to-machine interaction over a network. In practical terms, that means Web Services are application capabilities that can be used as standalone software components in a network [9]. These standalone capabilities can subsequently be linked or combined together without the need to understand the specific programming language, platform or application that is providing the information or functionality required. Web Services are explicitly defined, self-contained, and do not depend on the context or state of other interfaces.

Web Services currently offer one of the preferred methods for implementing Service Oriented Architecture (SOA) principles:
- Services are autonomous with well-defined inter-

faces.
- Services can be reused to build composite variations.
- Services are loosely coupled.
- Services are discoverable.
- Services are highly interoperable.

The key idea of SOA is the following: a service provider publishes services in a service registry. The service requester searches for a service in the registry. He finds one or more by browsing or querying the registry. The service requester uses the service description to bind service. These ideas are shown in the following Fig. 1.

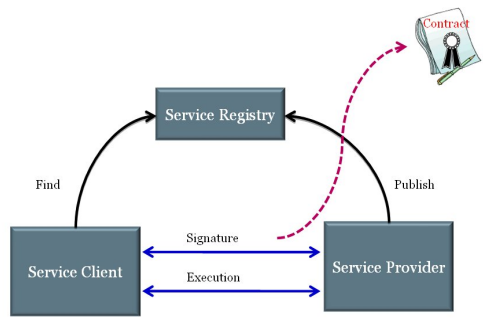

Fig. 1. Reference architecture of web services- SOA

The goal of Service Delivery Environement is to transform rigid network infrastructure to a standard, open, flexible, IP, SOA-based environment that can accommodate:
- Multiple traffic types (voice, data, and video).
- A multi-tiered service creation environment including IPTV, Content Delivery, Messaging, Legacy.
- Variety of access types (wired and wireless).
- Various Quality of Service (QoS) performance requirements.
- A variety of new and emerging business models.

Web Services can be used not only to interwork applications from multiple domains, but also to facilitate the integration and communication among operations and business support systems.

The Service Delivery Environement uses Web Services in combination with orchestration to implement processes such as fulfillment and assurance, and to expose charging and other services. Web Services are not the only means of exposing network and services abstractions and capabilities. Protocol-layer abstraction technologies such as IMS SIP/SCIM are also required for real-time and latency-intolerant services. For example within the telcom domain, SIP and Web Services must co-exist: they provide complementary ways of delivering context-aware, personalized and blended services.

**2.2 A Telemedicine Service Scenario**

A way to motivate and illustrate this work, we presents an example of Telemedicine service scenario. We categorize the Distributed Telemedicine environement into three main layers: service, business and delivery. The service layer consists of available web services, and the business layer represents the Web service like operations typically ordred in a particular application domain. We refer to the selected services as member services (see Fig. 2).

Key Telemedicine environement objectives include:
- Allowing people to stay in their homes to an older age. By doing this, we can reduce the economic burden of dedicated care facilities and improve quality of life for a substantial proportion of the aging population.
- Using televisions to keep in touch. Another use of camera technology is in conjunction with an IPTV set-top box and connection back to a Contact Center.
- Using wireless toys for always-on monitoring and communications. The wireless home network itself enables a new class of device that has significant healthcare implications.

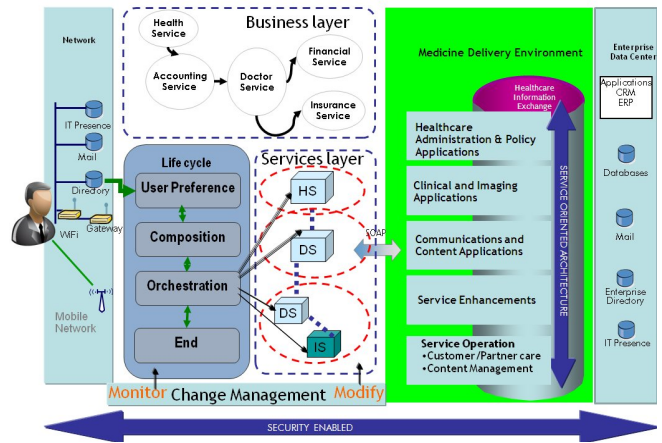

Fig. 2. Muti-tiered telemedicine application with different levels (Business, Services and Delivery)

Let us assume that a senior citizen establishes a need for a business objective (medicine service). Typically, he starts with formulating the business strategy (or goal) of the enterprise. During the planning, a set of services are identified: HealthService, AccountingService, DoctorService, FinancialServcie and InsuranceService. Second, the senior citizen develops a specification listing the services to be composed. A graphical interface is provided to senior citizen for easier management of the process. We assume that HS, AS, DS, FS and IS are selected and orchestrated. The third step is the orchestration where member services that match the specified high level configuration are selected and invoked. We describe here the ideal scenario: the senior citizen measures at home her blood sugar level, her blood pressure and her weight, and then he sends such data to the HealthService. Once the HealthService received the data, it controls whether the sender is a subscribed patient, and then it forwards the information who contacts it and when to AccountingService. HealthService forwards also the received data to DoctorService in charge of checking the received values. After analysing the received values, the team sends a confirmation or an adjustment of the medication doses. The FinancialService and InsuranceService are executed to provide finalise the process.

However some changes to member services may handle some inconsistency in the composition and orchestration. Each service layer change describes a functional and non functional change that may occur in a member service. For example, in case of non disponibility of a DoctorService, a change management is required to ensure that the telemedicine system is remaining profitable.

## 3 CHANGE SPECIFICATION

Managing buttom-up changes is highly dependent on the services that compose the system. Therefore, it is quite important to define the changes that may occur to web services and then map them to into the system level. In this section, we present buttom-up changes. We define the set of handling changes. Handling change ($\theta$) is defined for changes that occurred at the service level (for example Web service availability) while adaptive changes ($\Omega$) are related to changes at the business level (for instance the selection of alternative service).

### 3.1 Changes Overview

In this paper, we focus changes in synchronous mode. In our scenario, we suppose that for instance DS service may not change its data types while the triggering change of unavailability is being processed. Another hypothesis states that for each change a transition will be associated between two states: precondition and postcondition. In our case study presented, a precondition for DS unavailability is that it was available and the postcondition is that it has become unavailable.

Handling changes will be modeled using Petri nets. Our classification of triggering changes is based on the traditional approaches from the elds of software engineering and workfow systems [8]. A handling change is initiated at the service level such as the operations, the access points, the availability, etc. Therefore, we can distinguish several handling changes based on Web service properties [10].

The Web service properties can be sorted into two categories: functional and non-functional. We can classify handling changes into the two categories, as shown in Fig. 3.

- Non-functional changes: we assume that the non-functional parameters represent the dependability and response aspects associated with a member service.

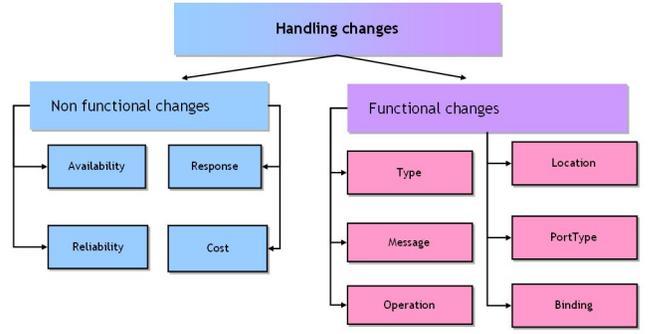

Fig. 3. Handling changes categories: functional and non-functional

dependability is associated with changes in the availability and reliability of the Web service. Service dependability takes only one of two possible values (i.e., available or unavailable). Alternatively, service cost values may take more than two possible values. For instance, the service cost obviously draws its values from the real domain. Changes are handled by values that exceed or fall below a threshold. This threshold consists of minimum and maximum values beyond which the needs to be notified about this change. Table 1 gives a summary about non-functional changes.

TABLE 1
NON-FUNCTIONAL CHANGES

| Change | Attribute | $\theta$ | Pre | Post |
|---|---|---|---|---|
| alterAvailibility | PSA | $\theta_A$ | PSA | PSA' |
| alterReliability | PSR | $\theta_R$ | PSR | PSR' |
| alterCost | PSC | $\theta_C$ | PSC | PSC' |
| alterResponsiveness | PSRe | $\theta_{Re}$ | PSRe | PSRe' |

- Functional changes: this category of changes is related to a service WSDL description [11]. We represent functional changes as a combined execution of elementary operations: remove and then add. We distinguish two different functional changes: structural and behavioral (see Fig. 3). Structural changes refer to the operational aspects of a Web service. For example, a structural change in a telemedicine service can be caused by changing the operations offered to a citizen. Changes to the behavior of a Web service are indicated by changing its interaction with external entities. Functional changes to a member Web service occur when its WSDL description is modified. Table 2 gives a summary about functional changes.

TABLE 2
FUNCTIONAL CHANGES

Remove

| Change | Attribute | $\theta$ | Pre | Post |
|---|---|---|---|---|
| removeType | PT | $\theta_T$ | PT | PT- |
| removeMessage | PM | $\theta_M$ | PM | PM- |
| removeOperation | PO | $\theta_O$ | PO | PO- |
| removeLocation | PL | $\theta_L$ | PL | PL- |
| removePortType | PPT | $\theta_{PT}$ | PPT | PPT- |
| removeBinding | PB | $\theta_B$ | PB | PB- |

Add

| Change | Attribute | $\theta$ | Pre | Post |
|---|---|---|---|---|
| addType | PT | $\theta_T$ | PT | PT+ |
| addMessage | PM | $\theta_M$ | PM | PM+ |
| addOperation | PO | $\theta_O$ | PO | PO+ |
| addLocation | PL | $\theta_L$ | PL | PL+ |
| addPortType | PPT | $\theta_{PT}$ | PPT | PPT+ |
| addBinding | PB | $\theta_B$ | PB | PB+ |

Adaptive changes: may occur at the composition and orchestration levels. In our scenario, when the telemedecine system is interrupted by a change in DS, it reacts to the change after suspending execution. This may be accomplished by raising a fault, compensating for the change at the composition layer, and calling of an alternate service. Each reaction may be explicitly executed by user as reactive or adaptive change or can be delegated to business layer which is considered as implicit reaction.

For example, if DS becomes unavailable, business layer will be search for equivalent service to continue execution to ensure that there is no high level impact on user demands. Table 3 gives a summary of adptive changes defined in our model.

TABLE 3
ADAPTIVE CHANGES

| Change | Attribute | $\Omega$ | Pre | Post |
|---|---|---|---|---|
| removeMember | VEM | $\Omega_M-$ | VEM | VEM- |
| addMember | VEM | $\Omega_S+$ | VEM | VEM+ |
| removeParameter | VEP | $\Omega_P-$ | VEP | VEP- |
| addParameter | VEM | $\Omega_S+$ | VEP | VEP+ |
| removeInstance | VEP | $\Omega_C-$ | VEC | VEP- |
| addInstance | VEC | $\Omega_C+$ | VEC | VEC-I |
| changeState | VES | $\Omega_S$ | VES | VEI-S |
| changeServiceInstance | VEI | $\Omega_I$ | VEI | VEI-SI |

Now, we will disucss the impact of $\theta$ changes to the telemedicine business layer. A mapping details how change instances in one layer correspond to changes in another layer. These mapping must remain consistant in the presence of frequent changes. Handling changes have a reactive impact on the business layer. For instance, a $\theta$ change in availability maps to $\Omega$ change of change instance.

# 4 CHANGE MODEL

In this section, we propose a change model to accurately identify all type of changes that may occur in a composite Web services.

## 4.1 Handling Changes Model Using Petri-Nets

Petri nets or PN are a well-founded process modeling techniques that have formal semantics. They have been used to model and analyze several types of processes including protocols, and business processes. Visual representations provide a high-level, yet precise language, which allows reasoning about concepts at their natural level of abstraction. Services are basically a partially ordered set of changes.

Therefore, it is a natural choice to map it into a Petri net. Moreover, the semantics delivered by Petri nets can be used to model the standard behavior of composite Web services described by BPEL [12].

We formalize the change model for triggering changes by introducing PN–Handle (PNH) which is defined as follows.

The algebric structure of PNH = (P, T, F, $P_0$, $P_n$) if the following conditions hold:
- F $\subseteq$ (P x T) $\cup$ (T x P)
- P $\cap$ T = $\varnothing$
- P $\cup$ T $\neq \varnothing$
- $P_i \in$ P
- $P_0 \in$ P

where:
- P is a finite set of places representing the states of Web service.
- F is a finite set of transitions representing changes to Web service.
- F is called the web services action flow.
- $P_0$ is the input place, or starting state of the Web service
- $P_n$ is the output place, or the ending state of the Web service

Fig. 4 represents the model of non-functional changes to Web services. It is composed from five places and four transitions. PS is the initial place of $PNH_N$. It represents the initial state of the Web Service. PS consists of four tokens, each representing one of the six non-functional changes. Every time a change occurs, the corresponding token is fired. If more than one change occurs, the corresponding token for each change type is fired.

For instance, if a member services (i.e Web service)

becomes unavailable, the transition will be enabled and the corresponding token will be fired.

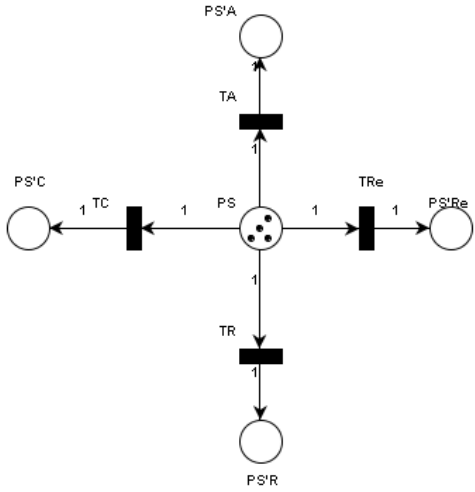

Fig. 4. Handling changes categories: functional and non-functional

The subnet representing dependability changes in $PNH_d = (P_d, F_d, T_d, P_{0d}, P_{nd})$, where $P_d=\{PS,PS\text{-}Re,PS\text{-}A\}$ and $F_d=\{FRe,FA\}$,. The place PS is corresponding to the state of available and reliable service. PS-A represents a service that becomes unavailable. When a service becomes unavailable, the token (representing the avalibility property) is moved from PS to PS-A. A similar behavior is observed when the service becomes unrelable. PS-A represents alterAvailibility, and PS-Re represents alterReliability. We notice that we have the same evolution of Petri nets model to represent a change in another non-functional property.

Fig. 5 represents the model of functional changes to Web service. Functional Changes are modeled by the subnet $PNH_f = (P_f, F_f, T_f, P_{0f}, P_{nf})$, where $P_f=\{P,PT+,PT-,PL+,PL-,PM+,PM-,PPT+,PPT-,PO+,PO-,PB+,PB-\}$ and $F_f=\{FT+,FT-,Fl+,FL-,Fm+,FM-,FPT+,FPT-,FO+,FO-,PB+,PB-\}$.

## 4.2 Modeling Adaptive changes with Reconfigurable Petri-Nets

We have surveyed several extensions of Petri nets for modeling reactive changes. Reconfigurable Petri-nets provide formalism for modeling these changes.

They support internal and incremental description of changes over a uniform description. Reconfigurable Petri nets are an extension of Petri nets with local structural modifying rules performing the replacement of one of its subnets by other subnets [13]. The tokens in a deleted place are transferred to the a created one.

We formalize the change model for adpative changes by introducing PNAC (PNAC) which is defined as follows. The algebric structure of PNAC= (P, T, F, R, I) where:

- $P=\{p_1,\ldots,p_n\}$ is a non empty and finite set of places
- $T=\{t_1,\ldots,t_n\}$ is a non empty and finite set of transitions disjoint from $P(P \bigcap T = \Phi)$
- F: $(P \times T) \bigcup (T \times P) \to$ IN is a weighted flow relation. A rewiting rule is a map $r: P_1 \to P_2$ whose domain and codomain are disjoint subsets of places $P, P_1 \subseteq P, P_1 \bigcap P = \Phi$
- $R=\{r_1,..,r_n\}$ is a finite set of structure modifying structure rules.
- I represent the initial state: the first configuration of composition in business layer. The domain of I is $DTE_0$.

We consider the scenario containing five places correspoding to adaptive changes:

- $DTE_S$ is the set of places $\{DTE_0, DTE_1, DTE_2, DTE_3, DTE_4\}$ where S represents changeState.
- $DTE_v$ is the set of places $\{DTE_5, DTE_6, DTE_7, DTE_8, DTE_9\}$ where V denotes changeServiceInstance.
- $DTE_w$ is the set of places $\{DTE_{10}, DTE_{11}, DTE_{12}, DTE_{13}, DTE_{14}\}$

Fig. 6 shows a PNAC representing adaptive changes used Reconfigurable Petri net. It represents the initial statechange in service orchestration, removal of service (Fig. 8), and addition of service (Fig. 9).

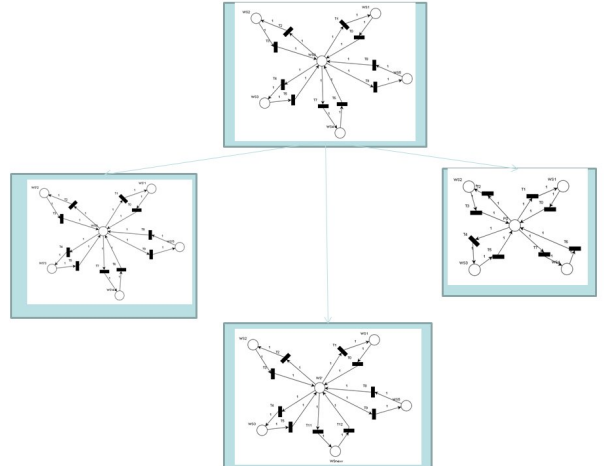

Fig. 6. PNAC for adaptive composition changes

Fig. 7 shows the PNAC after change in service orchestration.

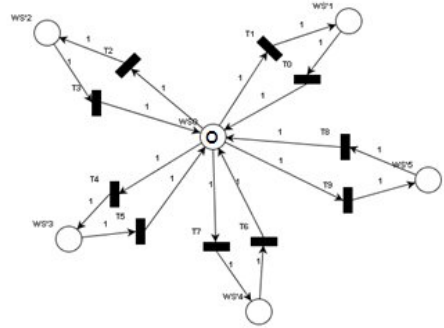

Fig. 7. PNAC initial statechange – the first configuration

The PNAC corresponding to removal of service is illu-strated by Fig. 8.

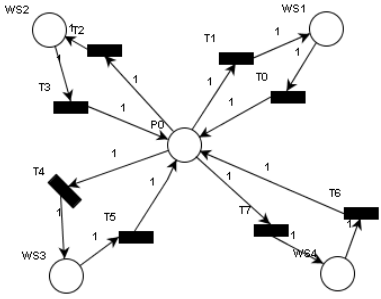

Fig. 8. PNAC after service removal- WS5 was removed

The last PNAC after of a new service is illustrated by Fig. 9.

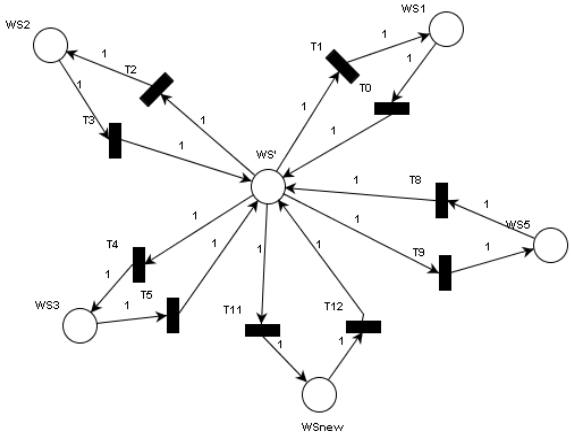

Fig. 9. PNAC after service addition- WSnew added

## 4.3 CHANGE MANAGEMENT APPORACH

Change management requires an automated approach that is specifically defined and automatically executed. We use our Petri net change specification as the basis for handling changes in our telemedicine environement. We divide the process of change management into two distinct steps: detection, and reaction.

After the change specification is defined, we begin the automated management. Detecting the respective changes is the first step of change management. All changes identified in the handling changes models are subject to detection. Detection involves an agent that monitors the Web service (Fig. 11). Each change type has an associated set of rules for detection. For example, a DoctorService may change the input parameters (i.e required information provided by patient), when this change occurs, the telemedicine system must detect this change using some predefined detection rules. Each detected change must be forwarded to monitoring service; and then the composition strategy must be updated. The notification and polling mechanism are mainly the techniques to awareness that a change has occurred. These techniques require that a monitoring service periodicaly send "refresh" and "Alive" messages to detect unavailable services and also renew membership. A Petri net based model of changes explained in section 5 represents the detection of handling changes. The changes are detected at the service layer and represented as an incidence matrix. Some rules are identified for detecting functional and non functional changes.

We define a rule for mapping the change into the defined Petri-net: first of all, the current service state is corresponded to a set of precondition places in the triggering Petri net and the updated service state as the set of postcondition places in the triggering Petri net. Then, a comparison between the values precondition and postcondition places of the Petri net is done. Depending on result retuned, a token is placed in the respective precondition place. This token will enable the change transition only if a difference is found.

Let us consider the example where the service WS service change availability (due to maintenance reasons) and the other attributes remain constant. In this case, we map the change into the non-functional Petri net (Fig. 10).

The service agent responsible of monitoring of WS will generate the following matrix detailed in Table 4.

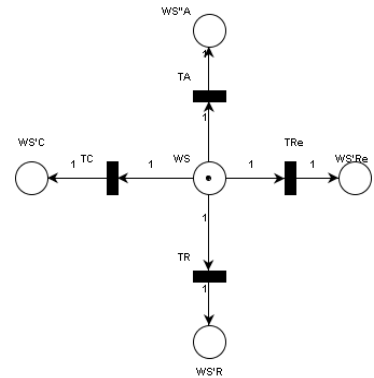

Fig. 10. Non-functional Petri net for WS- change of availibility of this service

TABLE 4
MATRIX OF NON-FUNCTIONAL CHANGE OF WS

|   | $\theta_A$ | $\theta_R$ | $\theta_C$ | $\theta_{Re}$ |
|---|---|---|---|---|
| WS'A | 1 | 0 | 0 | 0 |
| WS'R | 0 | 0 | 0 | 0 |
| WS'C | 0 | 0 | 0 | 0 |
| WS'R | 0 | 0 | 0 | 0 |
| WS | -1 | 0 | 0 | 0 |

Based on the information sent by service agent we define how to execute adaptive change. After receiving the matrix indicating the change that occurred, the handling change is mapped to the appropriate reactive change. We can list some considered reaction techniques in our system:

- In case of add, the newly service member will be considered. It can be considered in load balancing context or as back-up alternative
- In case of unavailability of a service, if it is critical then the orchestration will be paused. A heartbeat is activated to check the status of the service; the orchestration must be terminated if the heartbeat number permitted is reached.

## 5 CONCLUSION

In this paper, we have presented a muti-tiered architecture of Telemedicine services as well as the bottom-up approach focusing on handling changes that may occur in this system and then mapped to adaptive changes. We use a formal change model based on Petri nets to accurately represent these changes.

Future work includes extending our change management approach. We plan to include a top-down approach to specifying changes. Future work includes a full simulation prototype taking into account priority in changes, estimation of frequency based on measures and enhancing the monitoring module by proptagating optimally the changes and addition of expert module based on pre-defined user rules depending on his intention and preference.

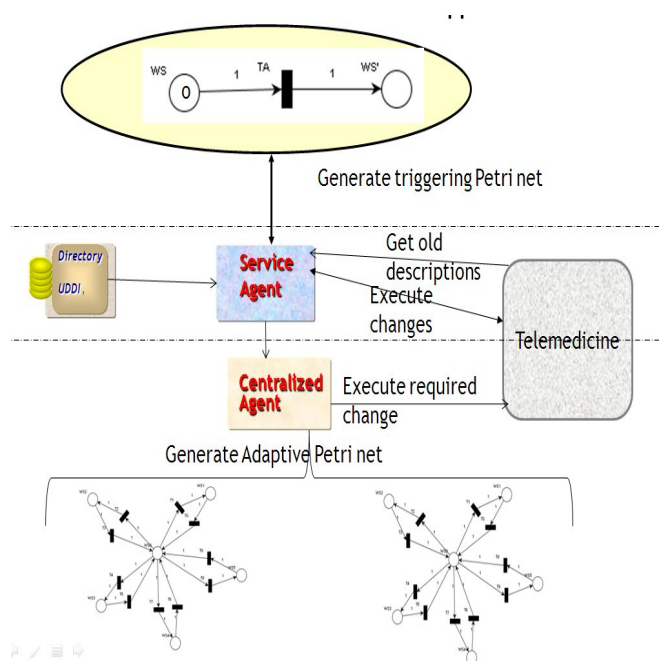

Fig. 11. Change management : main steps followed to generate the Adaptive Petri net

**S. Mtibaa** is currently a Ph.D. student in the National School for Computer Sciences of Tunis, Tunisia (ENSI). He received the master degree from High School of Communication of Tunis, University of Carthage, Tunisia (Sup'Com) in 2008. His current research interest includes web service composition using Petri nets as well as system verification and QoS aware.

**M. Tagina** is professor of Computer Science at the National School for Computer Sciences of Tunis, Tunisia (ENSI). He received the Ph.D. in Industrial Computer Science from Central School of Lille, France, in 1995. He heads research activities at LI3 Laboratory in Tunisia (Laboratoire d'Ingénierie Informatique Intelligente) on Meta-heuristics, Diagnostic, Production, Scheduling and Robotics.